\documentstyle[aps,epsfig,pra]{revtex}

\newcommand\beq{\begin{equation}}
\newcommand\eeq{\end{equation}}
\newcommand\bea{\begin{eqnarray}}
\newcommand\eea{\end{eqnarray}}

\def\eps{\epsilon}
\newcommand{\ket}[1]{| #1 \rangle}
\newcommand{\bra}[1]{\langle #1 |}

\newcommand{\QED}{\Box}
\newcommand{\ba}{\begin{array}}
\newcommand{\ea}{\end{array}}
\newtheorem{theo}{Theorem}
\newtheorem{defi}{Definition}
\newtheorem{lem}{Lemma}
\newtheorem{exam}{Example}

\newtheorem{cor}{Corollary}

\renewcommand{\>}{\rangle}
\newcommand{\<}{\langle}

\begin{document}

\title{The entanglement of purification}

\author{Barbara M. Terhal$^{1,2}$, Micha\l{} Horodecki$^3$, Debbie W. Leung$^2$ and David~P.~DiVincenzo$^{1,2}$}

\address
{$^1$ Institute for Quantum Information, California Institute of Technology, Pasadena, CA 91125, USA  \\
$^2$ IBM T.J. Watson Research Center, P.O. Box 218, Yorktown Heights, NY 10598, USA \\
$^3$ Institute of Theoretical Physics and Astrophysics,
University of Gda\'nsk, 80--952 Gda\'nsk, Poland}

\maketitle
\begin{abstract}
We introduce a measure of both quantum as well as classical correlations in a 
quantum state, the entanglement of purification. We show that the (regularized) entanglement of purification is equal to the entanglement cost of creating a state $\rho$ asymptotically from maximally entangled states, with negligible communication. We prove that the classical mutual information and the quantum mutual information divided by two are lower bounds for the regularized entanglement of purification. We present numerical results of the entanglement of purification for Werner states in ${\cal H}_2 \otimes {\cal H}_2$. 
\end{abstract}


\section{Introduction}

The theory of quantum entanglement aims at quantifying and characterizing uniquely quantum correlations. It does so by analyzing how entangled quantum states can be 
processed and transformed by quantum operations. A crucial role in the theory is played by the class of Local Operations and Classical Communication (LOCC), since quantum entanglement is non-increasing under these operations. 
Indeed, by considering this class of operations we are able to neatly distinguish between the quantum entanglement and the classical correlations that are present in the quantum state. 

Given the success of this theory, we may be daring enough to ask whether we can similarly construct a theory of purely classical correlations in quantum states and their behavior under local or nonlocal processing. At first sight, such an effort seems doomed to fail since merely local actions can convert quantum entanglement into classical correlations. Namely, Alice and Bob who possess an entangled state $\ket{\psi}=\sum_i \sqrt{\lambda_i} \ket{a_i} \otimes \ket{b_i}$ with Schmidt coefficients $\lambda_i$ can, by local measurements, obtain a joint probability distribution with mutual information equal to $H(\lambda)$. Thus it does not seem possible to separate the classical correlations from the entanglement if we try to do this in an operational way. Note that it may be possible to separate quantum and classical correlations in a nonoperational way, see for example Ref. \cite{ved_henderson} or \cite{cerf&adami}. The drawback of such an approach is that no connection is made to the dynamical processing of quantum information, which is precisely what has made the theory of quantum entanglement so elegant and innovative. An operational approach to the quantification of quantum and classical correlations was recently formulated in Ref. \cite{opphor}.

In this paper we propose to treat quantum entanglement and classical correlation in a unified framework, namely we express both correlations in units of entanglement. Such a theory of 'all' correlations may have potential applications outside quantum information theory as well. Researchers have started to look at entanglement properties of many-particle systems for example at (quantum) phase transitions (see for example Ref. \cite{osborne_nielsen:pt} and references therein). Instead of considering the entanglement of formation in these studies, one may choose to look at the behavior of a complete correlation measure. In this paper we introduce such a measure, called the entanglement of purification. We would like to emphasize that our correlation measure is {\em not} an entanglement measure, but a measure of correlations expressed in terms of the entanglement of a pure state.

It has been the experience in (quantum) information theory that
questions in the asymptotic approximate regime are easier to answer
than exact non-asymptotic queries. Thus we ask how to create a bipartite quantum state $\rho$ in the asymptotic regime, allowing approximation, from an initial supply of EPR-pairs by means of {\em local operations and asymptotically vanishing  communication}. This latter class of operations will be denoted as LOq (Local Operations with o(n) communication in the asymptotic regime) versus the class LO for strictly Local Operations. We can properly define this formation cost $E_{LOq}$ as follows:
\beq
E_{LOq}(\rho)=\lim_{\epsilon\rightarrow 0}\,\inf \left\{{m\over n}\,|\,
\exists\, {\cal L}_{LOq}, \right. \, \left.
D({\cal L}_{LOq}(\ket{\Psi_-}\bra{\Psi_-}^{\otimes m}),\rho^{\otimes n}) 
\leq \eps \right\}.
\label{defent}
\eeq
Here $\ket{\Psi_-}$ is the singlet state in ${\cal H}_2 \otimes
{\cal H}_2$ and ${\cal L}_{LOq}$ is a local superoperator using $o(n)$ quantum communication. $D$ is the Bures distance $D(\rho,\rho')=2\sqrt{1-F(\rho,\rho')}$ and the square-root-fidelity is defined as $F(\rho,\rho')={\rm Tr}(\sqrt{\rho^{1/2}\rho' \rho^{1/2}})$ \cite{uhlmann_fid}. We could have allowed classical instead of quantum communication in our definition, --our results will not depend on this choice--, so we may as well call all communication quantum communication.

Before we consider this entanglement cost for mixed states, we observe that by allowing asymptotically vanishing communication, we have preserved the interconvertibility result for pure states \cite{bbps}. This is due to the fact that both the process of entanglement dilution as well as entanglement concentration can be accomplished with no more than asymptotically vanishing amount of communication, see Ref. \cite{lopop}.

We see that the cost $E_{LOq}(\rho)$ of creating the state $\rho$ is defined analogously to the entanglement cost $E_c(\rho)$ \cite{bdsw}, \cite{hht:cost}, with the restriction that Alice and Bob can only do a negligible amount of communication. It is immediate that $E_{LOq}(\rho)$ will in general be larger than $E_c(\rho)$. In particular, for a separable density matrix $E_c(\rho)=0$ whereas we will show that for any correlated (i.e. not of the form $\rho_{AB}=\rho_A \otimes \rho_B$) density matrix  $E_{LOq}(\rho) > 0$. 
The entanglement cost $E_c$ was found \cite{hht:cost} to be equal to \beq
E_c(\rho)=\lim_{n \rightarrow \infty} \frac{E_f(\rho^{\otimes n})}{n},
\eeq
 where $E_f(\rho)$ is the entanglement of formation \cite{bdsw}. We will similarly find an expression for $E_{LOq}$
\beq
E_{LOq}=\lim_{n \rightarrow \infty} \frac{E_p(\rho^{\otimes n})}{n} \equiv E_p^{\infty}(\rho),
\eeq
where $E_p(\rho)$ is a new quantity, the entanglement of purification of $\rho$. 

Our paper is organized in the following manner. We start by defining the entanglement of purification and deriving some basic properties of this new function, such as continuity and monotonicity under local operations. We will relate the entanglement of purification to the problem of minimizing the entropy of a state under a local TCP (Trace-preserving Completely Positive) map. With these tools in hand, we can prove our main result, Theorem \ref{maintheo}. Then we spend some time proving the mutual information lower bounds for $E_{LOq}(\rho)$. We also compare our correlation measure with the induced Holevo correlation measures $C_{A/B}$ that were introduced in Ref. \cite{ved_henderson}. We prove that for Bell-diagonal states the correlation measure $C_A$ is equal to the classical capacity of the related 1-qubit Pauli channel. At the end of the paper we present our numerical results for $E_p(\rho)$ where $\rho$ is a Werner state on ${\cal H}_2 \otimes {\cal H}_2$. 
The proofs of the lemmas and theorems in this paper are all fairly straightforward and use many basic properties of entropy and mutual information (concavity, subadditivity of entropy, nonincrease of mutual information under local actions etc.).

\section{Entanglement of Purification}

We define the entanglement of purification: 

\begin{defi}
Let $\rho$ be a bipartite density matrix on ${\cal H}_A \otimes {\cal H}_B$.
Let $\ket{\psi} \in {\cal H}_{AA'} \otimes {\cal H}_{BB'}$. The
entanglement of purification $E_{p}(\rho)$ is defined as
\beq
E_p(\rho)=\min_{\psi: {\rm Tr}_{A'B'}\ket{\psi}\bra{\psi}=\rho}
E_f(\ket{\psi}\bra{\psi}),
\label{defep}
\eeq 
\end{defi}
where $E_f(\ket{\psi}\bra{\psi})$ is the entanglement of $\ket{\psi}$ which is equal to the von Neumann entropy $S(\sigma_{BB'})=-{\rm Tr} \sigma_{BB'}\log \sigma_{BB'}$ where $\sigma_{BB'}={\rm Tr}_{AA'}\ket{\psi}\bra{\psi}$.
Let $\{ \lambda_i,\ket{\psi_i}\}$ be the eigenvalues and eigenvectors
of $\rho_{AB}$.  The ``standard purification'' of $\rho$ is defined as 
\beq 
\ket{\psi_{\rm s}} = 
\sum_i \sqrt{\lambda_i} \ket{{\psi_i}}_{AB} \otimes \ket{0}_{A'} \ket{i}_{B'}.
\label{sp}
\eeq 
Every purification of $\rho$ can be written as $|\psi\> = (I_{AB}
\otimes U_{A'B'}) |\psi_{\rm s}\>$ for some unitary operator 
$U_{A'B'}$ on $A'$ and $B'$.  Therefore, Eq. (\ref{defep}) can be
rephrased as:
\bea
	E_p(\rho) & = & \min_{U_{A'B'}} E((I_{AB} \otimes U_{A'B'})
	\ket{\psi_{\rm s}}\bra{\psi_{\rm s}}
	(I_{AB} \otimes U_{A'B'})^\dagger).
\label{defepU}
\\ 
	& = & \min_{U_{A'B'}} S( {\rm Tr}_{AA'}  
	(I_{AB} \otimes U_{A'B'})
	\ket{\psi_{\rm s}}\bra{\psi_{\rm s}}
	(I_{AB} \otimes U_{A'B'})^\dagger) ) 
\nonumber
\\
	& = & \min_{\Lambda_{B'}} S(
	(I_{B} \otimes  \Lambda_{B'}) (\mu_{BB'}(\rho))) \,, 
\label{TCP}
\eea 
where we have taken the trace over $A$ and $A'$ to obtain 
Eq. (\ref{TCP}), 
\beq
	\mu_{BB'}(\rho) = {\rm Tr}_{AA'} |\psi_{\rm s}\rangle 
					  \langle \psi_{\rm s}| \,, 
\label{defmu}
\eeq
and $\Lambda_{B'}(\nu) \equiv {\rm Tr}_{A'} U_{A'B'} ( 
\nu_{B'} \otimes \ket{0} \bra{0}_{A'} ) U_{A'B'}^\dagger$. The minimization in Eq. (\ref{TCP}) is over all possible TCP maps $\Lambda_{B'}$ since every TCP map can be implemented by performing a unitary transformation on the system and some ancilla and tracing over the ancilla.  
Note that the minimizations over $U_{A'B'}$ and $\Lambda_{B'}$ 
are equivalent.
Equations (\ref{defepU}) and (\ref{TCP}) provide two different
formulations of the same minimization.  Conceptually the first
formulation is based on purifications of $\rho$ and variation over
$U_{A'B'}$.  The second formulation is based on extensions of $\rho$,
$\sigma_{ABB'}$, such that ${\rm Tr}_{B'} \sigma_{ABB'} = \rho_{AB}$, 
and variation over $\Lambda_{B'}(\nu)$.
Both formulations will be used throughout the paper.

The idea of bipartite purifications was considered in
Ref. \cite{boudabuzek} where the authors proved that every correlated
state has, in our language, a nonzero entanglement of purification. 
If we would have included mixed states in the minimization 
in Eq. (\ref{defep}) and used the entanglement of formation as the entanglement measure, then the defined quantity would be equal to the entanglement of formation of $\rho$, since the optimal extension of $\rho$ is $\rho$ itself.

We put some simple bounds on $E_p(\rho)$. 
Intuitively, `the amount of quantum correlation in a state is smaller
than or equal to the total amount of correlation', or $E_f(\rho) \leq E_p(\rho)$.  
To prove this lower bound, let $\ket{\psi_{\rho}}=\sum_{i,j} \ket{i}_{A'}
\ket{j}_{B'} \otimes \ket{\psi_{ij}}$ be the purification that
achieves the minimum in Eq. (\ref{defep}).  Alice and Bob locally
measure the labels $i_{A'}$ and $j_{B'}$ of the state
$\ket{\psi_{\rho}}$ such that they obtain $\ket{\psi_{ij}}$ with
probability $p_{ij}=\bra{\psi_{ij}} \psi_{ij} \rangle$.  Since
entanglement is nonincreasing under local operations, we have
\beq
E_f(\rho) \leq \sum_{ij} p_{ij}
E\left(\frac{\ket{\psi_{ij}}\bra{\psi_{ij}}}{p_{ij}}\right) \leq
E_p(\rho).
\eeq
It is immediate that we have equality between the entanglement of
formation and the entanglement of purification for pure states, where
the optimal purification of a pure state is the pure state
itself. 
%
%

An easy upper bound is $E_p(\rho) \leq E(\ket{\psi_{\rm s}}\bra{\psi_s}) =
S(\rho_A)$, where $\rho_A = {\rm Tr}_B (\rho)$ is the reduced density
matrix in $A$.  This corresponds to $U_{A'B'} = I_{A'B'}$ or
equivalently $\Lambda_{B'} = I_{B'}$ in the r.h.s. of
Eq. (\ref{defepU}) or (\ref{TCP}).
Applying the same argument with $AA'$ and $BB'$ interchanged, we obtain
\beq
E_p(\rho) \leq \min(S(\rho_A),S(\rho_B)),
\label{enbound}
\eeq
where the purifications correspond to either completely purifying the
state on $A'$ or on $B'$.  In general this is not the optimal 
purification, as we will see in Section \ref{wernerst}.

The entanglement of purification is neither convex nor concave, unlike
the entanglement of formation.
For instance, a mixture of product states, each with zero entanglement
of purification, need not have zero entanglement of purification (for
example, consider an equal mixture of $\ket{00}$ and $\ket{11}$).
On the other hand, the completely mixed state has zero entanglement of
purification equal to zero yet it is a mixture of 4 Bell states, each
with 1 ebit of entanglement of purification.  
%

Before we present continuity bounds for the entanglement of
purification, we analyze the optimization problem of Eq. (\ref{defep})
in more detail.  We can omit doubly stochastic maps $\Lambda_{B'}$ in
the optimization in Eq. (\ref{TCP}) since they never decrease the 
entropy.  
Furthermore, the von Neumann entropy is concave, so that the optimum
in Eq. (\ref{TCP}) can always be achieved when $\Lambda_{B'}$ is an
{\em extremal} TCP map. An extremal TCP map is a TCP map that cannot be expressed as a convex combination of other TCP maps. Choi \cite{choi75a} has proved that an extremal TCP map with input dimension $d$ has at most $d$ operation elements in its operator-sum representation. 
%
This result will allow us to upper bound the dimensions of the optimal
purifying Hilbert spaces, as stated in the following
Lemma.

\begin{lem}
Let $\rho$ act on a Hilbert space of dimension $d_{AB}=d_A d_B$. The
minimum of Eq. (\ref{defep}) can always be achieved by a state $\psi$
for which the dimension of $A'$ is $d_{A'}=d_{AB}$ and the dimension
of $B'$ is $d_{B'}=d_{AB}^2$ (or vice versa).
\label{optnum}
\end{lem}

{\em Proof}: We use the formulation of the entanglement of
purification as an optimization of a TCP map in Eq. (\ref{TCP}).
Since the density matrix $\mu_{BB'}(\rho)$ is on ${\cal H}_{d_B}
\otimes {\cal H}_{d_{AB}}$, the optimal map $\Lambda_{B'}$ maps ${\cal
H}_{d_{AB}}$ into a space of some unspecified dimension.  The
optimal map $\Lambda_{B'}$ can be assumed to be extremal.  Theorem 5 of
Choi \cite{choi75a} shows that an extremal TCP map $\Lambda: B({\cal
H}_{d_1}) \rightarrow B({\cal H}_{d_2})$\footnote{We have a special
case when $d_2=\infty$. The Stinespring theorem \cite{stinespring}
implies that we have an operator-sum representation of such a
map. Then Choi's results on extremality apply, bounding the number of
operation elements, from which the final result can be proved.} can be
written with at most $d_1$ operations elements, that is, has the form
\begin{equation}
\Lambda(\rho)=\sum_{i=1}^{d_1}V_i\rho V_i^\dagger.\label{Kraus}
\end{equation}
In our case $d_1=d_{AB}$.  Consider implementing the TCP map by
applying a unitary operation $U$ to the input state with an ancilla
appended.  In our case, this ancilla can be taken as Alice's purifying
system $A'$, and $U$ acts on $A'B'$.  The dimension of the ancilla
$A'$ can always be taken to be the number of operation elements.  Thus
we have $d_{A'}=d_{AB}$.  The $B'$ dimension is equal to the output
dimension $d_2$ of the optimal map $\Lambda$, which is unconstrained
by the extremality condition.  However, we note that the operator
$\Lambda(\rho)$ of Eq. (\ref{Kraus}) has a rank of at most $d_{AB}^2$.
This is obtained by observing that the range of this operator is exactly
that of the vectors given by all the columns of the matrices $V_i$ for
all $i$ (the $V_i$ matrices have $d_1$ columns and $d_2$ rows).  Thus,
there exists a unitary operator $U$ that permits the construction of a
new map $\Lambda'=U\Lambda$ whose output is confined to the first
$d_1^2$ dimensions of the output space.  The operator $U$ may be
obtained explicitly via a Gram-Schmidt procedure applied to the column
vectors of the $V_i$ matrices.  $\Lambda'$ is also optimal, since the
entropy of Eq. (\ref{TCP}) is not changed by a unitary operation.
Since the output space of $\Lambda'$ has dimension $d_1^2$, we
conclude that $d_{B'}$ can be taken to be $d_{B'}=d_{AB}^2$. $\QED$

It is interesting to note that a similar minimization problem was encountered in
Ref. \cite{superdense_noisy}. There the goal was to use a set of noisy
states for classical information transmission and we wanted to minimize the coherent information divided by the entropy of a quantum state under the action of a local map.

\begin{theo}[Continuity of the Entanglement of Purification]
Let $\rho$ and $\sigma$ be two density matrices on ${\cal H}_{d_A} \otimes {\cal H}_{d_B}$ with Bures distance $D(\rho,\sigma) \leq \eps$. Then
\beq
|E_p(\rho)-E_p(\sigma)| \leq 20 D(\rho,\sigma) \log d_{AB}  -D(\rho,\sigma) \log D(\rho,\sigma),
\label{contep}
\eeq
for small enough $\eps$.
\label{conttheo}
\end{theo}

{\em Proof}: Let $\ket{\psi_{\sigma}'}$ and $\ket{\psi_{\rho}'}$ be the purifications 
of $\rho$ and $\sigma$ which achieve the maximum \cite{uhlmann_fid} in 
\beq
F(\rho,\sigma)=\max_{\psi_{\sigma},\psi_{\rho}} |\langle \psi_{\sigma} |
\psi_{\rho} \rangle|.
\eeq
Let $\ket{\phi_{\rho}}$ and $\ket{\phi_{\sigma}}$ correspond to the 
optimal purifications of $\rho$ and $\sigma$ with respect to $E_p$. There exists a unitary transformation $U$ relating $\ket{\psi_{\rho}'}$ to $\ket{\phi_{\rho}}$, i.e. $(U \otimes {\bf 1})\ket{\psi_{\rho}'}=\ket{\phi_{\rho}}$. We define the (non-optimal) purification $\ket{\psi_{\sigma}}$ as $(U \otimes {\bf 1}) \ket{\psi_{\sigma}'}=\ket{\psi_{\sigma}}$. Now we have 
\beq
E_p(\sigma)-E_p(\rho)=E(\ket{\phi_{\sigma}}\bra{\phi_{\sigma}})-E(\ket{\phi_{\rho}}\bra{\phi_{\rho}}) \leq E(\ket{\psi_{\sigma}}\bra{\psi_{\sigma}})-E(\ket{\phi_{\rho}}\bra{\phi_{\rho}}).
\eeq
We use continuity of entanglement \cite{bst,nielsen:cont}, Lemma \ref{optnum} (which indicates that the pure state has support on a space of dimension at most $d_{AB}^4$), and the fact that $|\bra{\psi_{\sigma}} \phi_{\rho} \rangle|=|\bra{\psi_{\sigma}'} \psi_{\rho}' \rangle|=F(\rho,\sigma)$ to bound 
\beq
E_p(\sigma)-E_p(\rho) \leq 5 D(\rho,\sigma) \log d_{AB}^4  -2 D(\rho,\sigma) \log D(\rho,\sigma).
\eeq
for small enough $D(\rho,\sigma)$. We can obtain the full bound in Eq. (\ref{contep}) by alternatively relating $\ket{\psi_{\sigma}'}$ to the optimal purification $\ket{\phi_{\sigma}}$ by a unitary transformation $U$. $\QED$

It is fairly straightforward to prove monotonicity of the entanglement of purification from monotonicity of entanglement:

\begin{lem}[Monotonicity of the Entanglement of Purification]
The entanglement of purification of a density matrix $\rho$ is nonincreasing under strictly local operations. Let Alice carry out a local TCP map ${\cal S}_A$ on the state $\rho$. We have 
\beq
E_p(({\cal S}_A\otimes {\bf 1})(\rho)) \leq E_p(\rho).
\label{sa}
\eeq
Let Alice carry out a local measurement on $\rho$ through which she obtains 
the state $\rho_i$ with probability $p_i$. We have
\beq
\sum_i p_i E_p(\rho_i) \leq E_p(\rho).
\eeq
Let ${\cal L}_{LOq}$ be a local operation assisted by $m$ qubits of communication.
The entanglement of purification obeys the equation
\beq
E_p({\cal L}_{LOq}(\rho)) \leq E_p(\rho)+m.
\eeq
\label{loinc}
\end{lem}

{\em Proof}: Let $\ket{\psi_{\rho}}$ be the optimal purification of $\rho$.
 This optimal purification is related to {\em some} purification of $({\cal S}_A \otimes {\bf 1})(\rho)$ by a unitary transformation on Alice's system only. Then Eq. (\ref{sa}) follows from the fact that entanglement is nonincreasing under local partial traces. The state $\ket{\psi_i}=\frac{A_i \otimes I_B \ket{\psi}}{\sqrt{\bra{\psi} A_i^{\dagger}A_i \otimes I_B \ket{\psi}}}$ where $A_i$ corresponds to a measurement outcome of Alice, is some purification of $\rho_i$. The entanglement is nonincreasing under local operations and thus
\beq
E_p(\rho)=E(\ket{\psi_{\rho}}\bra{\psi_{\rho}}) \geq \sum_i p_i E(\ket{\psi_i}\bra{\psi_i}) \geq \sum_i p_i E_p(\rho_i).
\eeq
For the last inequality, let Alice and Bob start with the entangled state $\ket{\psi_{\rho}}$ and carry out their LOq protocol. By subadditivity of entropy, the entanglement of this state can increase by at most $m$ bits when $m$ qubits of communication are sent (back and forth). Thus the entanglement of the final state which is some purification of ${\cal L}_{LOq}(\rho)$ is smaller than or equal to $E_p(\rho)+m$. $\QED$

Now we are ready to prove our main theorem:

\begin{theo}
The entanglement cost of $\rho$ on ${\cal H}_d \otimes {\cal H}_d$ without classical communication equals
$E_{LOq}(\rho)=E_p^{\infty}(\rho)$.
\label{maintheo}
\end{theo}

{\em Proof}: The inequality $E_{LOq}(\rho) \leq E_p^{\infty}(\rho)$ uses entanglement dilution. Let $k$ be the number of copies of $\rho$ for which 
the regularized entanglement of purification $E_p^{\infty}$ is achieved. One way of 
making many ($p$) copies of $\rho^{\otimes k}$ out of EPR pairs 
and $o(p) \leq o(pk)$ classical communication, is to first perform entanglement dilution on the EPR pairs so as to create (an approximation to) 
the purification $\ket{\psi}^{\otimes p}$ and then trace over the additional registers to get $\rho^{\otimes kp}$. The other inequality $E_p^{\infty}(\rho) \leq E_{LOq}(\rho)$ can be proved from monotonicity and continuity of the entanglement of purification. We start with $n$ EPR pairs which have $E_p$ equal to $n$. The LOq process for creating an approximation $\tilde{\rho}_k$ to $\rho^{\otimes k}$ using $o(k)$ qubits of communication, increases the entanglement of purification by at most $o(k)$ bits, see Lemma \ref{loinc}, or $E_p(\tilde{\rho}_k) \leq n+o(k)$. Using the continuity of Theorem \ref{conttheo} and dividing the last inequality by $k$ and taking the limit $k \rightarrow \infty$ gives $E_p^{\infty}(\rho) \leq E_{LOq}(\rho)$. $\Box$

\section{Mutual Information Lower bounds}

The entanglement cost $E_{LOq}$ is a measure of the quantum and classical correlations in a quantum state. The quantum and classical mutual information of a quantum state are similar measures that capture correlations in a quantum state. How do these measures relate to the new correlation measure? The quantum mutual information $I_q(\rho_{AB})$ is defined as
\beq
I_q(\rho_{AB})=S(\rho_A)+S(\rho_B)-S(\rho_{AB}).
\label{qmi}
\eeq
We define the classical mutual information of a quantum state $I_c(\rho_{AB})$ as 
\beq
I_c(\rho_{AB})=\max_{M_A:p_A, M_B:p_B} H(p_A)+H(p_B)-H(p_{AB}).  
\label{defic}
\eeq
Here local measurements $M_A$ and $M_B$ give rise to local probability distributions $p_A$ and $p_B$. The classical mutual information of a quantum state is the maximum classical mutual information that can be obtained by local measurements by Alice and Bob.
Both quantum as well as classical mutual information share the important property that they are non-increasing under local operations (LO) by Alice and Bob.
For the classical mutual information, this basically follows from the definition Eq. (\ref{defic}). The definition itself as a maximum over local measurements makes sense since the classical mutual information of a probability distribution is 
non-increasing under local manipulations of the distribution. The proof of this well known fact is analogous to the proof for the quantum mutual information which we will give here for completeness.

We can write the quantum mutual information as
\beq
I_q(\rho_{AB})=S(\rho_{AB}||\rho_A \otimes \rho_B),
\eeq
where $S(.||.)$ is the relative entropy. The relative entropy is nonincreasing under any map $\Lambda$ (cf. Ref. \cite{vedral&plenio:entanglement}), i.e.
\beq
S(\Lambda(\rho_{AB})||\Lambda(\rho_A \otimes \rho_B)) \leq S(\rho_{AB}||\rho_A \otimes \rho_B).
\eeq
When $\Lambda$ is of a local form, i.e. $\Lambda_A \otimes \Lambda_B$, the l.h.s. of this equation equals the quantum mutual information of the state $(\Lambda_A \otimes \Lambda_B)(\rho_{AB})$ and thus the inequality $I_q((\Lambda_A \otimes \Lambda_B)(\rho_{AB})) \leq I_q(\rho_{AB})$ is proved.

\subsection{Proof of Lower bounds}

We show that the quantities $I_q(\rho)/2$ and the regularized classical information $I_c^{\infty}(\rho)=\lim_{n \rightarrow \infty}\frac{I_c(\rho^{\otimes n})}{n}$ are both lower bounds for the entanglement cost $E_{LOq}$. The argument is similar to the proof of the $E_p^{\infty}$ lower bound on $E_{LOq}$ in Theorem \ref{maintheo} (The reasoning is in fact a special case of Theorem 4 in Ref. \cite{mhorodecki:measures} (cf. Ref. \cite{dhr:unique}) applied to the class LOq instead of the original LOCC.) 

We start with a number, say $k$, of EPR pairs which have $I_q=2k$ and $I_c$ equal to $k$ \footnote{One can prove that $I_c \leq k$ by observing that any local measurement that is not projecting in the Schmidt basis is a noisy version of the measurement that does project in the Schmidt basis. In other words, the probability distribution of any set of local measurements can be obtained from the probability distribution of the Schmidt basis measurement by local processing, which does not increase the classical mutual information. \label{foot1}}. In the limit of large $n$ , the ratio $k/n$ is the entanglement cost $E_{LOq}(\rho)$. We apply the LOq map ${\cal L}$ which uses $o(n)$ communication to obtain an approximation ${\tilde \rho}_n$ to $\rho^{\otimes n}$. Since the quantum mutual information and the classical mutual information can only increase by $o(n)$ by the process ${\cal L}$ applied to the initial EPR pairs, see Lemma \ref{noincrease}, it follows that 
\beq
I_q({\tilde \rho}_n) \leq o(n)+2k, 
\label{b1}
\eeq
and similarly
\beq
I_c({\tilde \rho}_n) \leq o(n)+k.
\label{b2}
\eeq
The last step is to relate the mutual informations of ${\tilde \rho}_n$ to 
the mutual informations of $\rho^{\otimes n}$. For this, we need a continuity result of the form 
\beq
|I_{q/c}(\sigma)-I_{q/c}(\rho)| \leq C \log d ||\rho-\sigma||_1 +O(1).
\label{contqc}
\eeq
for $\rho, \sigma$ on ${\cal H}_d$, $||\rho-\sigma||_1$ sufficiently small and $C$ is some constant \footnote{We can alternatively write down a continuity relation using the Bures distance. Since the trace-distance $||.||_1$ and the Bures distance are equivalent distances, one continuity relation follows from the other and vice versa.}.
Below we will prove these desired continuity results.
We can divide Eqs. (\ref{b1}) and (\ref{b2}) by $n$ and take the limit of large $n$. We use the continuity relation of Eq. (\ref{contqc}) and the fact that in the large $n$ limit ${\tilde \rho}_n$ tends to $\rho^{\otimes n}$. Thus we have 
\beq
\lim_{n \rightarrow \infty} \frac{I_q(\rho^{\otimes n})}{n}=I_q(\rho) \leq 2E_{LOq}(\rho), 
\eeq
where we used that the quantum mutual information is additive, and similarly
\beq
I_c^{\infty}(\rho) \leq E_{LOq}(\rho).
\eeq

What remains is to prove the continuity relations and the nonincrease modulo $o(n)$ under LOq operations. 

\subsubsection{Continuity of Mutual Informations}

The continuity of the quantum mutual information $I_q(\rho)$ can be proved by invoking Fannes' inequality \cite{fannes} and Ruskai's proof of nonincrease of the trace-distance under TCP maps \cite{ruskai}. Let $\rho$ and $\sigma $ be two density matrices which are close, i.e. $||\rho-\sigma||_1={\rm Tr}|\rho-\sigma| \leq \eps$ for sufficiently small $\eps$. We have
\beq
|I_q(\rho_{AB})-I_q(\sigma_{AB})| \leq |S(\rho_A)-S(\sigma_A)|+|S(\rho_B)-S(\sigma_B)|+|S(\sigma_{AB})-S(\rho_{AB})|, 
\eeq
which can be bounded as
\beq
|I_q(\rho_{AB})-I_q(\sigma_{AB})| \leq 3 \log d_{AB} || \rho_{AB}-\sigma_{AB}||_1+3 \eta(||\rho_{AB}-\sigma_{AB}||_1),
\eeq
where $\eta(x)=-x\log x$ and $||\rho-\sigma||_1 \leq 1/3$.

It is not hard to prove the continuity of the classical information of a quantum state, again using the nonincrease of $||.||_1$ under TCP maps. Let $M_A^{\rho}$ and $M_B^{\rho}$ be the optimal measurement achieving the classical mutual information $I_c(\rho)$. Under this measurement the states $\rho$ and $\sigma$, which is, say, close to $\rho$, go to probability distributions $p^{\rho}(i,j)$ and $p^{\sigma}(i,j)$ which are close again, i.e. $||p^{\rho}-p^{\sigma}||_1 \leq ||\rho-\sigma||_1$. We have that 
\beq
I_c(\sigma)-I_c(\rho) \leq I(p^{\sigma})-I(p^{\rho}) \leq \log k ||p^{\rho}-p^{\sigma}||_1+O(1),
\label{classfannes}
\eeq
where $k$ is the number of joint outcomes in the optimal measurement $(M_A^{\rho},M_B^{\rho})$ and $I$ is the classical mutual information of a joint probability distribution. The last inequality in Eq. (\ref{classfannes}) could in principle be derived from Fannes' inequality, using diagonal matrices, but it is a standard continuity result in information theory \cite{cover&thomas:infoth} as well. To finish the argument, we should argue that $k$, the number of joint measurement outcomes is bounded. The classical mutual information $I$ is a concave function of the joint probability $p(i,j)$ \cite{cover&thomas:infoth}. Therefore only extremal measurements $M_A$ and $M_B$ need to be considered in the optimization over measurements. An extremal measurement has at most $d^2$ outcomes when acting on a space of dimension $d$ \cite{peresbook} and thus $k \leq d_{AB}^2$. The same argument, interchanging $\sigma$ and $\rho$, can be used to upperbound $I_c(\rho)-I_c(\sigma)$.

\begin{lem}[Monotonicity Properties of Mutual Information]
Let ${\cal L}$ consist of a series of local operations assisted by $m$ qubits of 2-way communication. The quantum mutual information obeys the inequality
\beq
I_q({\cal L}(\sigma)) \leq I_q(\sigma)+2m,
\label{iqin}
\eeq
for all states $\sigma$. For the classical mutual information we have 
\beq
I_c({\cal L}(\ket{\psi}\bra{\psi}) \leq I_c(\ket{\psi}\bra{\psi})+m,
\label{icin}
\eeq
for all pure states $\ket{\psi}$.

\label{noincrease}
\end{lem}

{\em Proof}:
Let us first consider the quantum mutual information.
We can decompose the 2-way scheme ${\cal L}$ into a sequence of one-way schemes. It is sufficient to prove for such a one-way scheme using $m$ qubits of communication, say from Alice to Bob, that
\beq
I_q({\cal L}(\sigma)) \leq I_q(\sigma)+2m.
\eeq
Alice's local action can consist of adding an ancilla $A'$ in some state and apply a TCP map to the systems $AA'$ thus obtaining the state $\sigma_{AA':B}$. Such an action does not increase the quantum nor classical mutual information as we showed before. Now Alice sends system $A'$ to Bob. We have 
\bea
\lefteqn{
I_q(\sigma_{AB})\geq I_q(\sigma_{AA':B})=S(AA')+S(B) - S(AA'B)\geq } \nonumber \\ 
& & S(AA')-S(A')+S(BA') - S(AA'B) \geq S(A)-2S(A')+S(BA')- S(AA'B) =I_q(\sigma_{A:BA'})-2S(A'),
\eea
where we used $|S(A)-S(B)| \leq S(AB) \leq S(A)+S(B)$. 
The quantum mutual information of the final state is 
$I_q(\sigma_{A:BA'})$. Since $S(A')\leq m$, we obtain the needed inequality. Alice could send only a part of ancilla $A'$, but this does not change the bound.

Let us now consider the classical mutual information. We may convert the entire process ${\cal L}$ into a coherent process ${\cal L}$ where all the measurements are deferred to the end, this does not change the amount of communication that Alice and Bob carry out. Thus, prior to the measurements Alice and Bob have converted the pure state $\ket{\psi}$ into some pure state $\ket{\phi}$ whose local entropy is at most $E+m$ where $E$ is the entanglement of the state $\ket{\psi}$, which is equal to $I_c(\ket{\psi}\bra{\psi})$ (see footnote \ref{foot1}). Now Alice and Bob locally measure and/or trace out some registers which are operations that do not increase the classical mutual information. Therefore the final state ${\cal L}(\ket{\psi}\bra{\psi})$ has a classical mutual information that is bounded by the initial classical mutual information plus $m$.
$\QED$.

{\em Remark}: Note that Eq. (\ref{iqin}) for the quantum mutual information
applies to both pure and mixed states while we have found mixed states
that violate Eq. (\ref{icin}) for the classical mutual information.

Let us state the final result once more: 
\begin{cor}
$E_{LOq}(\rho)\geq I_q(\rho)/2$ and $E_{LOq} \geq I_c^{\infty}(\rho).$
\end{cor}
 
With this Corollary we can show that the LOq-entanglement cost of any correlated density matrix $\rho$, is nonzero \footnote{Note that this does not directly follow from the result in Ref. \cite{boudabuzek}, since the entanglement of purification may be nonadditive.}. Indeed, the quantum mutual information $I_q(\rho)$ of a correlated density matrix is strictly larger than zero, since $S(\rho_{AB})$ is strictly less than $S(\rho_A)+S(\rho_B)$ (equality is only obtained when $\rho_{AB}=\rho_A \otimes \rho_B$) and therefore $E_{LOq}(\rho) > 0$.

We present a simple example for which $E_{LOq}(\rho)=E_p^{\infty}(\rho) > I_q(\rho)/2$.
\begin{exam}[All correlation is classical correlation]
Consider the separable state $\rho=\sum_i p_i \ket{a_i}\bra{a_i}\otimes \ket{b_i}\bra{b_i}$ where $\bra{a_i}a_j\rangle=\delta_{ij}$ and $\bra{b_i}b_j\rangle=\delta_{ij}$. In this case $I_q(\rho)/2=H(p)/2$. However we can show that 
$E_p(\rho) \geq H(p)$. We have (cf. Eq. (\ref{defmu})) $\mu(\rho)=\sum_i p_i \ket{b_i}\bra{b_i}\otimes \ket{i}\bra{i}$. Under some local TCP map $\Lambda$ we obtain a state $\mu'=\sum_i p_i \ket{b_i}\bra{b_i} \otimes \rho_i$ where $\rho_i$ are density matrices. The entropy of $\mu'$ equals $S(\mu')=\sum_i p_i S(\rho_i)+H(p)\geq H(p)$.
The entanglement of purification $E_p(\rho)$ may be nonadditive, so we have to consider $E_p(\rho^{\otimes n})$. We have $\mu(\rho^{\otimes n})=\mu^{\otimes n}$ and now $\mu'=\sum_{i_1,\ldots,i_n} p_{i_1}\ldots p_{i_n} \ket{i_1\ldots,i_n}\bra{i_1,\ldots,i_n}\otimes \rho_{i_1,\ldots,i_n}$. Again the von Neumann entropy of $\mu'$ is larger than or equal to $n H(p)$. Note that in this example we do achieve the classical mutual information lower bound.
\end{exam}

Here is an example where the upper and lower bounds fix the (regularized) entanglement of purification:
\begin{exam}
Let $\rho$ be an equal mixture of the state $\ket{\Psi_0}=\frac{1}{\sqrt{2}}(\ket{00}+\ket{11})$ and $\ket{\Psi_1}=\frac{1}{\sqrt{2}}(\ket{00}-\ket{11})$. Alice and Bob can get 1 bit of classical mutual information by both measuring in the $\{0,1\}$ basis. Thus $E_{LOq}(\rho) \geq I_{c}(\rho)=1$, but $E_{LOq}(\rho) \leq S(\rho_A) \leq 1$, Eq. (\ref{enbound}). Therefore $E_{LOq}=1$.
\end{exam}


\section{Other Correlation Measures: The locally Induced Holevo Information}

In Ref.\cite{ved_henderson} the authors considered the locally induced Holevo information as a measure of classical correlations in the state. It is defined either with respect to Alice's measurement ($C_A$) or Bob's measurement ($C_B$)
\beq
C_{A/B}(\rho)=\max_{M_A/M_B} S(\sum_i p_i^{B/A} \rho_i^{B/A})-\sum_i p_i^{B/A} S(\rho_i^{B/A}),
\label{inhol}
\eeq
where $M_A$ ($M_B$) on $\rho$ gives reduced density matrices $\rho_i^{B}$ ($\rho_i^A$) with probability $p_i^B$ ($p_i^A$). The classical mutual information $I_c^{\infty}(\rho)$ will in general be less than these quantities, since to achieve the Holevo information one may have to do coding. In Ref. \cite{ved_henderson} it was shown that $C_{A/B}$ are nonincreasing under local operations. We leave it as an exercise for the reader to prove continuity and nonincrease modulo o(n) under LOq operations (applied to some pure state), thus showing that the regularized versions of these two quantities are also lower bounds for $E_{LOq}$.

\subsection{Bell-diagonal states}
\label{locin}

We show that for Bell-diagonal states $\rho_{Bell}$ the quantity $C_A$ (equal to $C_B$ by symmetry of the Bell-diagonal states) is equal to the classical capacity of the corresponding qubit channels. By the previous arguments this give us some lower bounds on the regularized entanglement of purification of these states. The Bell-diagonal states are of the following form
\beq
\rho_{Bell}=\sum_ip_i |\Psi_i\>\< \Psi_i|,
\eeq
where $\Psi_{0\ldots 3}$ are the four Bell states where $\ket{\Psi_0}$ is $\frac{1}{\sqrt{2}}(\ket{00}+\ket{11})$.
The corresponding channel, --the so called generalized depolarizing channel--, or Pauli channel, is of the form 
\beq
\Lambda_\rho(\cdot) = \sum_i p_i\sigma_i (\cdot)\sigma_i,
\label{unital}
\eeq
where $\sigma_0={\bf 1}$, and $\sigma_{1,2,3}$ are the three Pauli matrices. It is known \cite{mrpra} that all two qubit states with maximally mixed subsystems are Bell-diagonal, up to a unitary transformation $U_A \otimes U_B$. 
From the isomorphism between states and channels \cite{jamiolkowski,choi75a,bdsw}, it follows that all unital channels are of the form  (\ref {unital}) [cf. \cite{kingruskai}], up to unitary transformations applied before and after the action of the channel. 
The classical 1-shot capacity of the quantum channel $\Lambda$ 
is given by \cite{schum:cap,holevo:cap}
\beq
C_1(\Lambda)=\sup_{\{q_i,\rho_i \}}  
\chi(\{\ q_i,\Lambda(\rho_i)\}),
\eeq
where $\chi$ is the Holevo function of the ensemble 
\beq
\chi(\{\ q_i,\rho_i\})= S(\sum_i q_i\rho_i)-
\sum_i q_iS(\rho_i).
\eeq
The optimal states $\rho_i$ that achieve the capacity $C_1$ are always pure states, moreover it can be shown \cite{kingruskai} that the ensemble $\{q_i,\ket{\psi_i}\}$ that achieves $C_1$ for unital 1-qubit channels satisfies 
\beq
\sum_i q_i|\psi_i\>\<\psi_i|={1\over 2} {\bf 1}.
\label{ensem}
\eeq

Let us argue that $C_A(\rho)=C_1(\Lambda)$ for a Bell-diagonal state $\rho_{Bell}=({\bf 1}_A \otimes \Lambda_{\rho})(\ket{\Psi_0}\bra{\Psi_0})$. Alice's POVM measurement on this state commutes with the channel $\Lambda_{\rho}$. By doing a measurement on $\ket{\Psi_0}$ she can create any pure-state-ensemble on system $B$, obeying the relation Eq. (\ref{ensem}). This ensemble is then sent through the channel $\Lambda_{\rho}$. If the ensemble is optimal for $C_1$, then its Holevo information $\chi$ equals $C_1$ and thus $C_A=C_1$. 


For unital 1-qubit channels $C_1$ is given by 
\cite{kingruskai,king}
\beq
C_1(\Lambda)= 1-\min_\psi S(\Lambda(|\psi\>\<\psi|)).
\label{capacity}
\eeq
We can perform the minization in the last inequality and we obtain the following formula for the capacity of a Pauli channel or the induced Holevo information of the Bell-diagonal states 
\beq
C_A(\rho_{Bell})=C_1(\Lambda_{\rho})= 1-H(1-\lambda),
\eeq
where $\lambda$ is the sum of the two largest probabilities $p_i$ and $H(.)$ is the binary entropy function $H(x)=-x\log x -(1-x)\log(1-x)$. For a two-qubit Werner states of the form 
\beq
\rho_W = e \ket{\Psi_0} \bra{\Psi_0}
                    + (1-e)/3 \sum_{i=1}^3 \ket{\Psi_i} \bra{\Psi_i},
\label{wer}
\eeq
we obtain
\bea
C_A=1-H\left({1+2e\over 3}\right) ~~ {\rm for}~e \in [{1 \over 4},1], \nonumber
  \\
  C_A=1-H\left({2-2e\over 3}\right) ~~ {\rm for}~e \in [0,{1 \over 4}].
\label{cabound}
\eea

It was shown by King \cite{king} that the classical capacity of unital 1-qubit channels is equal to the one shot capacity, or $C_1=C_1^{\infty} \equiv \lim_{n \rightarrow \infty} \frac{1}{n} C_1(\rho^{\otimes n})$. Therefore $C_A=C_A^{\infty}=C_1$, which is a lower bound on $E_{LOq}$.

\section{Werner states}
\label{wernerst}

\begin{figure}
\begin{center}
\epsfxsize=8cm 
\epsffile{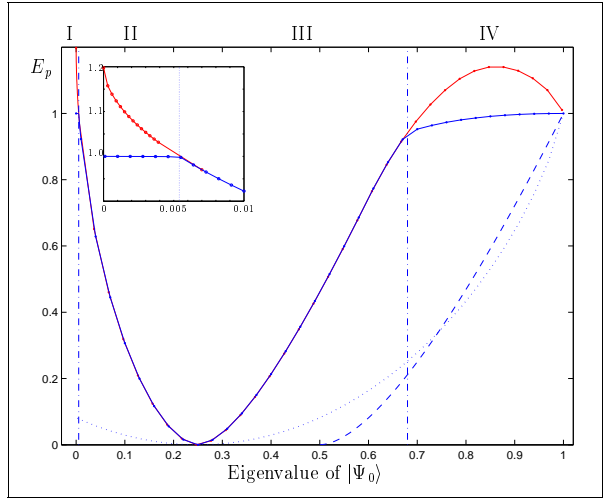}
\caption{Numerical bounds on $E_p$ for Werner states. In the upper
curve we restrict to ${\rm dim}(A') = {\rm dim}(B') = 2$; for the
next curve, we permit ${\rm dim}(A') = {\rm dim}(B') = 4$. The inset shows the curious behavior of $E_p$ around the point where the eigenvalue of $\ket{\Psi_0}$ approaches zero. The dotted
curve is the $C_A$ lower bound of Sec. \protect\ref{locin}. The dashed curve is the entanglement of formation lower bound which vanishes when the eigenvalue is smaller than or equal to $1/2$.}.
\label{fig1}
\end{center}
\end{figure}

A numerical minimization based on Eq. (\ref{defepU}) was performed for the
Werner states Eq. (\ref{wer}) for $E_p$.  We plot the results as a function
of the $\ket{\Psi_0}$ eigenvalue $e$ in Fig. \ref{fig1}.
We permitted
various output dimensions;  The two curves shown have ${\rm
dim}(A') = {\rm dim}(B') = 2$ and ${\rm dim}(A') = {\rm dim}(B') = 4$.
In the first case, the initial variable of the minimization was
determined by a random $4 \times 4$ unitary $U_{A'B'}$ picked
according to the Haar measure.
In the second case, the initial point was determined by a random 
$16 \times 4$ isometry picked according to a parameterization derived 
from Ref.~\cite{reck}.  We did not explore the largest dimensions
permitted by Lemma \ref{optnum}, which would have required an optimization
over a $64 \times 4$ isometry. 

It is evident from the numerics presented in the figure that the $C_A$
bound of Eq. (\ref{cabound}) is not achieved for the Werner states: the $C_A$ lower bound is only tight at the trivial points $e=1/4$ and $e=1$.
Our results indicate that $E_p$ is a very complex function, neither
concave nor convex, with several distinct regimes.  In fact, we find
four different regimes in our numerics: I) In this regime the standard
purification of Eq. (\ref{sp}) appears to be optimal, so the $U$ of
Eq. (\ref{defepU}) is the identity, and the purifying dimensions are
${\rm dim}(A') = 1$ and ${\rm dim}(B') = 4$.  This regime only extends
over a tiny range, approximately $0\leq e\leq 0.005$.  II) In the
range $0.005\leq e \leq 0.25$ we find an optimal purification of the
form 
\beq 
\sqrt{e} \ket{{\Psi_0}}_{AB}
\ket{{\Psi_0}}_{A'B'}+\sqrt{1-e\over 3}\left( \ket{{\Psi_1}}_{AB}
\ket{{\Psi_1}}_{A'B'}+ \ket{{\Psi_2}}_{AB} \ket{{\Psi_2}}_{A'B'}+
\ket{{\Psi_3}}_{AB} \ket{{\Psi_3}}_{A'B'}\right). 
\eeq 
In this region the $E_p$ curve is given by $E_p=-x\log x-(1-x)\log((1-x)/3)$, with $x=(1+2e-2\sqrt{3}\sqrt{e(1-e)})/12$.  Here the purifying dimensions
are ${\rm dim}(A') = 2$ and ${\rm dim}(B') = 2$.  Of course $E_p$
drops to zero for the completely mixed state at $e=1/4$.  III) In the
range $0.25\leq e\leq 0.69$ we also find purifying dimensions
${\rm dim}(A') = 2$ and ${\rm dim}(B') = 2$, but we were unable to
determine the analytical form of the purifying state or of $E_p$.
IV) In the range $0.69\leq e\leq 1$ the purifying dimensions were
${\rm dim}(A') = 2$ and ${\rm dim}(B') = 3$.  Again, we were unable
to come to any analytical understanding of the result.  Of course, $E_p=1$
for $e=1$, corresponding to the pure maximally entangled state. 

\section{Conclusion}
We have shown that the entanglement cost $E_{LOq}(\rho)$ is equal to the regularized entanglement of purification. It is an open question whether the entanglement of purification is additive:
\beq
E_p(\rho \otimes \rho)\stackrel{?}{=}E_p(\rho)+E_p(\rho).
\eeq 
In the alternative formulation using the state $\mu(\rho)$ the additivity question is the following. Is the minimum in 
\beq
\min_{\Lambda_{CD}} S((I_{AB} \otimes \Lambda_{CD})(\mu_{AC} \otimes \mu_{BD}),
\eeq
achieved by a TCP map $\Lambda_{CD}={\cal S} \otimes {\cal S}$? This problem is similar again to the additivity question encountered in Ref. \cite{superdense_noisy} where a local map could possibly lower the ratio of the coherent information and the entropy of many copies of a state together.

It is interesting not only to ask the formation question with respect to this class LOq, but also consider `the distillation' question. One can consider different versions. For example, how much entanglement can we distill from $\rho$ using $o(n)$ communication? One would expect that this quantity $D_{LOq}(\rho)$ is always zero for states for which the entanglement cost $E_c$ (using $LOCC$) is lower than the distillable entanglement $D$. We do not have a proof of this statement, relating irreversibility to a need for classical communication.

Instead of trying to convert the correlations in $\rho$ back to entanglement, we may ask what classical correlations Alice and Bob can establish using $\rho$.
We could allow Alice and Bob to perform an asymptotically vanishing amount of communication in this extraction process. A little bit of communication could potentially increase the classical mutual information in a quantum state by a large amount (when the classical correlation is initially 'hidden'), thus this may not be the best problem to pose. Researchers \cite{gisin_boundc} \cite{ww:unpublished} have investigated the possibly more interesting problem of the {\em secret} key $K$ that Alice and Bob can establish given $\rho$ where one allows arbitrary public classical communication between the parties. There is again more than one version of this problem, one in which Eve possesses the purification of the density matrix \cite{gisin_boundc} and a situation in which Eve is initially uncorrelated with the density matrix.  In Ref. \cite{massar_secret} a general framework is developed to address these issues also in the multipartite setting. 


Quite recently, entanglement properties of bipartite density matrices were studied by looking at mixed extensions of the density matrix \cite{tucci}. It would be interesting to explore the connection between our results here on the entanglement of purification and this other approach.

\section{Acknowledgments}
B.M.T., D.W.L. and D.P.D. are grateful for the support of the National Security Agency and the Advanced Research and Development Activity through Army Research Office contract numbers DAAG55-98-C-0041 and DAAD19-01-C-0056 and partial support from the National Reconnaissance Office. This work was also supported in part by the National Science Foundation under Grant. No. EIA-0086038. M.H. acknowledges the support of EU grant EQUIP, Contract No. IST-1999-11053. We thank Charles Bennett, Pawe\l{} Horodecki, Ryszard Horodecki and John Smolin for a pleasant IBM lunch discussion on this topic. The concept of entanglement of purification was raised in a discussion of M.H. with Ryszard Horodecki. M.H. would also like to thank Robert Alicki and Ryszard Horodecki for stimulating discussions. B.M.T would like to thank Andreas Winter for interesting discussions about the secret key rate $K$ and its relation to other correlation measures. D.W.L. would like to thank Charles Bennett and John Smolin for discussions on mixed state inputs that violate Eq. (\ref{icin}).   
\bibliographystyle{hunsrt}
\bibliography{refs}

\end{document}